\documentclass[useAMS,usenatbib]{mnras}
\usepackage[dvips]{graphicx}
\usepackage{amssymb,amsmath}
\usepackage{color}
\usepackage{epstopdf}
\usepackage{natbib}
\usepackage{lipsum}

\newcommand{\da}{the }

\title[Observable effects of ram pressure]{Following the crumbs: Statistical effects of Ram Pressure in Galaxies}
\author[Rodr\'iguez, Garcia Lambas, Padilla and Troncoso-Iribarren]{S. Rodr\'iguez$^1$ \thanks{E-mail:srrodriguez@famaf.unc.edu.ar}, D. Garcia Lambas$^{1,2}$, N. D. Padilla$^{3,4}$ and P. Troncoso-Iribarren$^5$\\
$^1$ Instituto de Astronom\'ia Te\'orica y Experimental, UNC-CONICET, C\'ordoba, X5000BGR, Argentina\\
$^2$ Observatorio Astron\'omico de C\'ordoba, Universidad Nacional de C\'ordoba, X5000BGR, Argentina\\
$^3$ Instituto de Astrof\'isica, Pontificia Universidad Cat\'olica de Chile, 8970117, Santiago, Chile\\
$^4$ Centro de Astro-Ingenier\'ia, Pontificia Universidad Cat\'olica de Chile, 8970117, Santiago, Chile\\
$^5$ Universidad Aut\'onoma de Chile, Chile. Av. Pedro de Valdivia 425, 7500912, Santiago, Chile}

\begin{document}

\date{Accepted --. Received --; in original form \today}

\pagerange{\pageref{firstpage}--\pageref{lastpage}} \pubyear{2019}

\maketitle

\label{firstpage}

\begin{abstract}

We analyse the presence of dust around galaxy group members through the reddening of background quasars. By taking into account quasar colour and their dependence on redshift and angular position, we derive mean quasar colours excess in projected regions around member galaxies and infer the associated dust mass. For disc-like galaxies perpendicular to the plane of the sky, and at group-centric distances of the order of the virial radius, thus likely to reside in the infall regions of groups, we find systematic colour excess values $e \sim 0.009 \pm 0.004$ for $g-r$ colour. Under the hypothesis of Milky Way dust properties we derive dust masses of $5.8 \pm 2.5 \cdot 10^8 M_{\sun}/h$, implying that a large fraction of dust is being stripped from galaxies in their path to groups.

We also studied the photometry of member galaxies to derive a colour asymmetry relative to the group centre direction from a given galaxy. We conclude that the regions of galaxies facing the centre are bluer, consistent with the effects of gas compression and star-formation.

We also combine these two procedures finding that galaxies with a small colour asymmetry show the largest amounts of dust towards the external regions compared to a control sample.
We conclude that dust removal is very efficient in galaxies on infall. The fact that galaxies redder towards groups centres are associated to the strongest reddening of background quasars suggest that gas removal induced by ram pressure stripping plays a key role in galaxy evolution and dust content.

\end{abstract}

\begin{keywords}
  galaxies: fundamental parameters -– intergalactic medium –- galaxies: groups: general -- galaxies: star formation -- quasars: general -- surveys
\end{keywords}

\section{Introduction}
\label{intro}

Galaxies have evolved simultaneously with their environment and at the present-day they have formed groups and clusters via their infall onto denser regions. This fact is key to explain galaxy properties in systems though the action of mergers \citep{barnes96}, ram pressure \citep{gunn72} and harassment \citep{moore96} that shape and affect galaxies as they fall onto a group or cluster \citep[for a review see][]{boselli06}. In particular, ram pressure stripping from the intra-cluster medium can play a key role in removing gas from galaxies, thus affecting future star-formation once the galaxies have entered in these systems. Besides the fact that properties of galaxies in clusters differ from those elsewhere, there is also mounting observational evidence of stripped material from galaxies in groups and clusters \citep{Poggianti2016, jaffe18}.

Numerical simulations also provide useful information on the transformation process of galaxies. For instance, hydrodynamical simulations show how efficiently ram pressure quenches star-formation during infall \citep[see for instance][]{steinhauser16}. Also, semi-analytical models need to model the effect of ram pressure to reproduce some observed galaxy properties \citep{delucia04, tecce10, guo11}. Ram pressure stripping is an effective mechanism by which a galaxy can lose a significant amount of gas and dust and so, galaxies in their first infall onto a group/cluster can suffer a strong change leaving a substantial fraction of gas and dust in their trajectories onto clusters.

\citet[hereafter T16]{TI16} studied the efficiency of star formation of the leading and trailing halves of galaxies according to the direction of the galaxy motion in their orbits around clusters in the EAGLE hydrodynamical simulation \citep{eagle1, eagle2}. The authors find a significant correlation between the SFR efficiency and the relative angle between the velocity vector and the group centre. Galaxies in infall, particularly at large relative angles, show an enhanced SFR efficiency in their leading halves with respect to the trailing one(see right panel of Fig. 2 in T16). We stress the fact that these results deal with 3-D analysis with no projection effects. In a companion work, T19 (in preparation), the authors also find that the SFR enhancement is mainly owed to an increase of the pressure and density of the gas particles in the leading half which boosts the star-formation.

In this work we search for different observational effects of environment in galaxies in groups and their neighbourhoods using data from the Sloan Digital Sky Survey \citep[SDSS,][]{sdss} Data Release 8 \citep{dr8} Two different effects driven by ram pressure stripping on galaxies will be studied. We will explore the presence of debris stripped from galaxies by means of the observed colour excess of background quasars \citep[following the work of][]{mcgee2010} around galaxies that are members of groups. Secondly, following T16 and T19, we will analyse traces of systematic differences in star-formation rates in different zones of galaxies and its possible relation to the stripped material. 

This work is organised as follows: In \S \ref{method} we explain in detail our procedures to measure the two effects mentioned in the previous paragraph. In \S \ref{sample} we describe the sample used in the analysis. In \S \ref{results} we show the results obtained, with the results of the colours of background quasars in \S \ref{results_crumbs}, the results of the analysis of the colours in different zones of galaxies in \S \ref{results_oranges}, and in \S \ref{results_both} we explore the connection between both effects. Finally in \S \ref{conclusions} we present our final remarks and conclusions.

\section{Methods}
\label{method}

\begin{figure}
	\begin{center}
		\includegraphics[width=.45\textwidth]{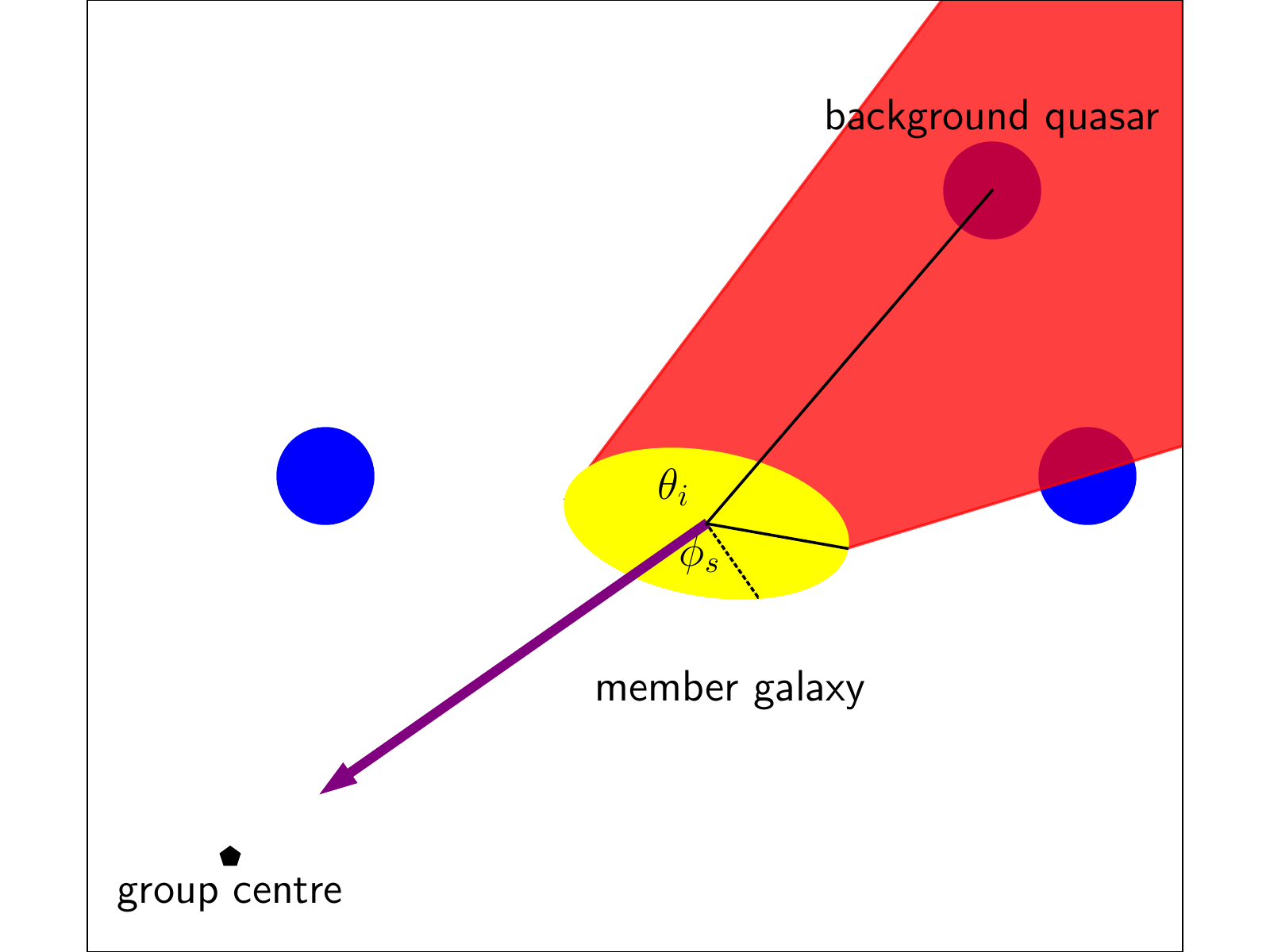}
		\caption{\label{fig_dia} A group galaxy (yellow) and background quasars (blue) projected in the sky plane. The dotted line is perpendicular to the vector between the galaxy and the centre of the group and separates the galaxy in two halves. $\phi_s$ is the angle between the semi-mayor axis of the galaxy and the vector to the centre of the group. $\theta_i$ is the angle between the vector from the centre of the group to the galaxy and the vector between the galaxy and the quasar. The red area represents the reddening by galaxy debris.}
	\end{center}
\end{figure}

In order to asses the presence of dust associated to member galaxies, we characterise background quasars excess colours in zones near to these galaxies.
For this aim, we select group member galaxies (with low semi-mayor/semi-minor axis ratio, $b/a < 0.5$), located in projection at least at 0.5 halo radius from the centre of their corresponding group. By doing so, we maximise the number of galaxies in the infall region \citep[see Fig. 6 in][]{jaffe18}. We consider a circle of 600 kpc$/h$ projected radius around each galaxy and select background quasars within these circles. We notice that a given quasar can be associated to more than one galaxy, in this case the quasar information is used more than once. For each quasar, we calculate the angle between the vector of the corresponding galaxy to the centre and the vector between the member galaxy and the quasar. Fig. \ref{fig_dia} schematises the position of the member galaxy, the group centre and the background quasar, and the angle $\theta_i$ which ranges between -180$^{\circ}$ and 180$^{\circ}$. With this setup we stack galaxies scaling to the angular size corresponding to 600 kpc$/h$ in projection and use $\theta_i$ as the angular position for each of the background galaxies, oriented with respect to the direction of the centre of each group.

Since it could be possible that the dependence of colour excess on $\theta_i$ is not due to the presence of galaxies but rather intrinsic to groups, we create control samples of positions within groups.
To construct these samples we take the position and the angular size of the projected 600 kpc$/h$ radius of each of the member galaxies and replace it by a random position at the same distance to the centre of the group with no other galaxy within 300 kpc$/h$ of projected separation. We obtain a sample of background quasars for these control zones. The difference of quasar colour excess between galaxy and control zones as a function of $\theta_i$ can then be considered as caused by the presence of debris associated to the member galaxies.

\subsection{Difference between halves.}
\label{crumbs}

Additionally, we provide an analysis of the impact of ram pressure by looking at colour asymmetries   of galaxies taken from the same sample of member galaxies studied in the previous case including those members at less than 0.5 halo radius from the centre. For each of these galaxies we use the images in the $r$ and $b$ bands, and we obtain photometry corresponding to an ellipse with semi-major axis $\sqrt{a/b}$ $r_{90}$ and semi-minor axis $\sqrt{b/a}$ $r_{90}$ that contain $90$ percent of the galaxy total flux in each band. The ellipses are then separated in two halves by a line perpendicular to the vector to the centre of the group. We remark the fact  that we cannot use the velocity vector as in T16 given the lack of kinematic information. In Fig. \ref{fig_dia} this line appears as a dashed line in the member galaxy. Following T16, we call the half ellipse closer to the centre as the leading-half, and the other one as the trailing-half. Then, we use the two bands to calculate the $g-r$  colour for each half and calculate the relative colour difference between the two halves. T16 and T19 define trailing/leading half the sides separated according to the three-dimensional velocity vector, while here we use the projected direction to the group centre.

The calculation of the relative difference of colour is done as follows: For each half we calculate the colour ($C_l$ for the leading, $C_t$ for trailing), then the difference $\Delta C = C_l-C_t$.

This allows a study of a possible relation between the values of $\Delta C$ and the angle between the semi-mayor axis and the vector to the centre (In Fig. \ref{fig_dia} labelled as $\phi_s$). Any correlation between the internal asymmetries and the intragroup dust mass content would be mostly interesting for the interpretation of the results.

\section{Group, galaxy and quasar samples}
\label{sample}

We use the group catalogue from \citet{yang2012}. This catalogue has been produced using DSS-DR7 \citep{dr7}. Since the \citet{yang2012} groups are identified assuming an overdensity of 180 we use $r_{180}$ as halo radius, computed using equation 5 in \citet{yang2007}. From the total group sample we consider only those with at least 4 member galaxies with a halo mass larger than $10^{12.5} M_{\sun}/h$. 

Additionally, we select galaxies brighter than -15 absolute magnitude in the $r$ band and with $r_{90}$ values, measured in the $r$-band, of at least 5 arcsec, and $b/a<0.5$. This last restriction is applied in order to restrict our sample to galaxies nearly edge-on so that if at infall onto the groups in the plane of the sky, their projected area of debris is larger than the area of face-on galaxies, since those galaxies would have their motion perpendicular to the disc planes.

We expect to find the most significant effects in late-type galaxies due to their higher amount of dust and cold gas with respect to early-type ones. In order to measure $\Delta C$ values for individual galaxies, we take galaxy images in the $g$ and $r$ bands reprocessed in SDSS-DR8 \citep{dr8}. The number of selected galaxies for this analysis is 12440, which reside in 5323 groups. Table \ref{tab_medians} shows the median values of magnitudes and colours for this galaxy sample.

\begin{table}
	\begin{center}
		\caption{\label{tab_medians} The median values of magnitudes and colours.}
		\begin{tabular}{lr}
								& Median 			  \\\hline
			$M_r - 5\log(h)$	& $-19.658 \pm 0.013$ \\
			$m_r$				& $16.782 \pm 0.009$  \\
			$m_g$				& $17.596 \pm 0.009$  \\
			$g-r$				& $0.87 \pm 0.002$    \\
		\end{tabular}
	\end{center}
\end{table}

The adopted sample of quasars to derive reddening in regions around group galaxies was taken from the  BOSS quasar sample \citep{BOSS}. These background quasars are further restricted to the redshift range $2<z<3.5$  since the maximum of the quasar distribution is in this region, and we aim to avoid possible incompleteness issues. The total number of quasars within the  area covered by groups is 139294.

\begin{figure*}
  \begin{center}
    \includegraphics[width=.4\textwidth]{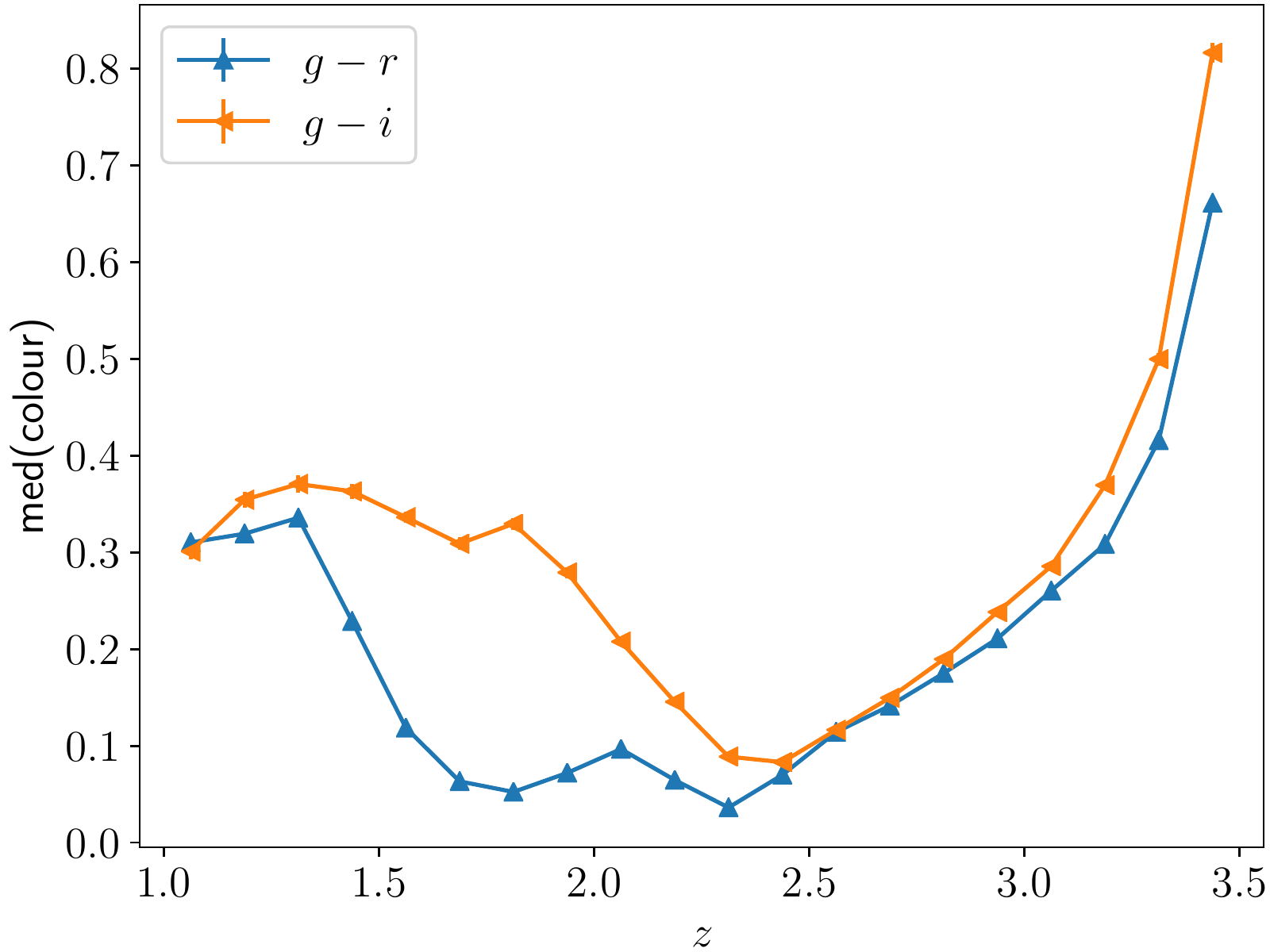}
    \includegraphics[width=.29\textwidth]{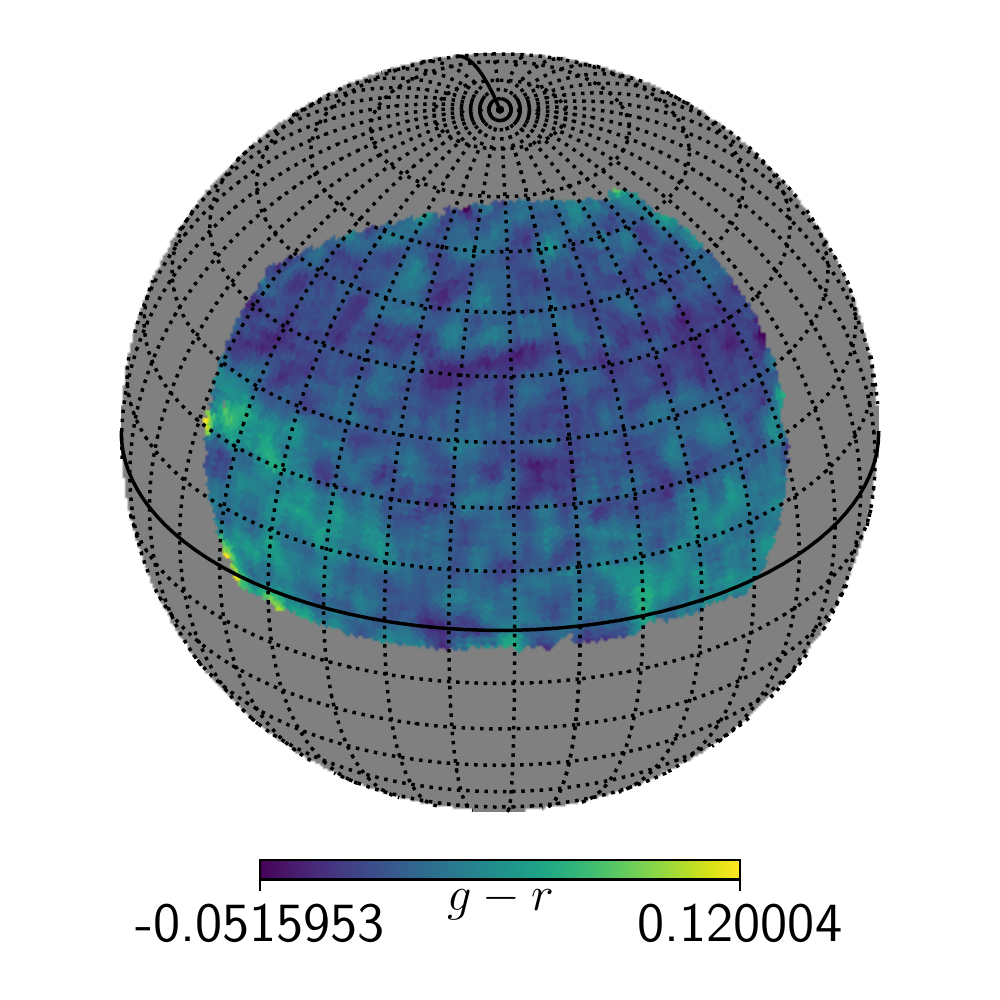}
    \includegraphics[width=.29\textwidth]{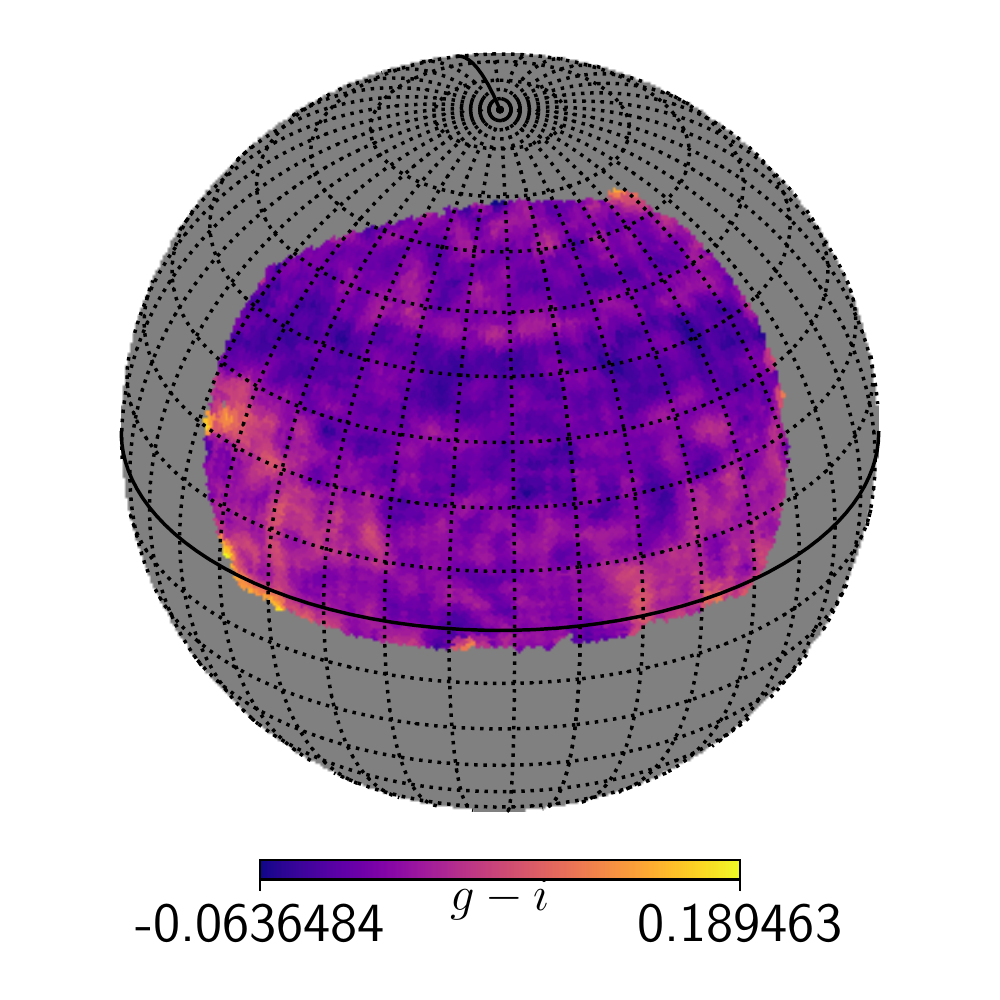}
    \caption{\label{fig_corrMaps} Corrections applied to the colour of quasars. LEFT: Dependence of $g-i$ and $g-r$ with redshift. MIDDLE: Extinction map for $g-r$ colour calculated using a 2 degrees low pass filter. RIGHT: Extinction map for $g-i$.}
  \end{center}
\end{figure*}

It is necessary to take into account several effects on the observed quasar colour besides the reddening from dust associated to local structures. The K-correction \citep{oke1968} is strongly redshift dependent and so as a first step we consider the relation between the median observed quasar colour and redshift (left panel of Fig. \ref{fig_corrMaps}). With this information we compute the mean excess colours of individual quasars taking into account the expected colour as a function of quasar redshift. This correction allows to decrease the colour dispersion by 20\%, which in turn improves our measurement uncertainties by 30\%. The other relevant  effect is due to the extinction and reddening from the interstellar \citep[see for instance][]{schlegel98} and zodiacal dust \citep{may07}. To tackle this issue, we measure the median of quasar excess colours (previously redshift corrected) according to their position in the sky. Then we apply a 2 degrees low pass filter to construct maps which are only affected by reddening at scales larger than 2 degrees (middle and right panels in Fig. \ref{fig_corrMaps}) since 2 degrees is larger than the largest 600 kpc area around galaxies in our sample. Thus, the total correction applied to quasar colours are the sum of the previously described redshift corrections plus the angular correction from the interstellar medium.

In order to minimise the effects of the presence of strong absorption regions close to the galactic plane we restrict our samples of  galaxies and quasars to galactic latitude $\lambda>40^\circ$.

\section{Results}
\label{results}

\subsection{Observational effects of the debris}
\label{results_crumbs}

\begin{figure*}
	\begin{center}
		\includegraphics[width=.45\textwidth]{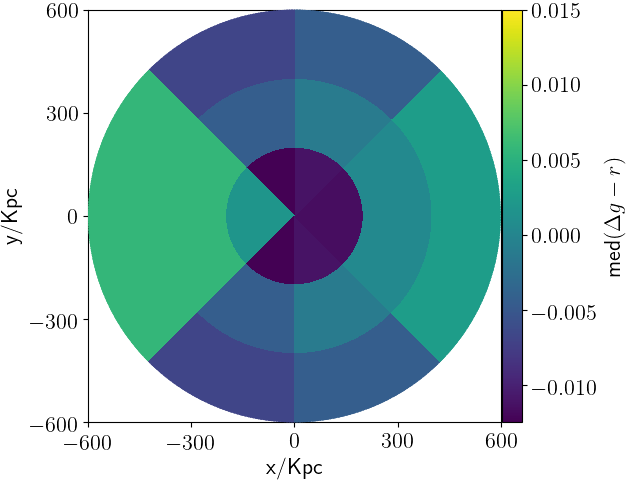}
		\includegraphics[width=.45\textwidth]{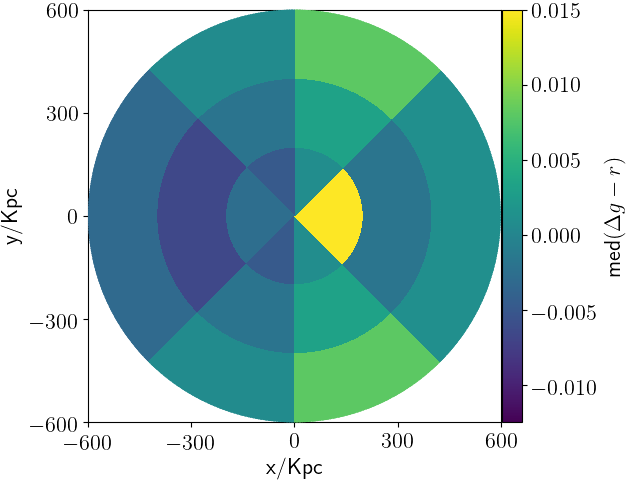}
		\caption{\label{fig_group_mat} Colour map of background quasars around galaxies, the group centre is the direction of the positive $x$ axis. In these maps we calculate $\theta_i$ in the range 0$^{\circ}$. - 180$^{\circ}$, and then mirror the data for the 180$^{\circ}$. - 360$^{\circ}$ range. LEFT: Maps for $g-r$ colour around galaxies. RIGHT: Maps for $g-r$ colour around control zones.}
	\end{center}
\end{figure*}

Fig. \ref{fig_group_mat} shows the angular distribution of the $g-r$ colour excess of background quasars associated to galaxies in groups. The group centre direction is to the right along the positive  $x$ axis. We can see a positive colour excess of background quasars as we move away form the group centre. However, a similar excess can be observed in the direction to the centre of the group, an effect that could be owed to the group medium and is not necessarily related to the presence of a galaxy. To explore this issue into more detail, we compare these results with those derived from the control sample where a similar  colour excess is seen in the direction to the group centre, while in the positive radial direction, colours of background quasars tend to be bluer.

A total of 10399 quasars were used for our analysis, and on average, 7 quasar sightlines crossed each galaxy. We find that approximately 26\% of quasars passed through more than one target galaxy.

\begin{figure*}
	\begin{center}
		\includegraphics[width=.45\textwidth]{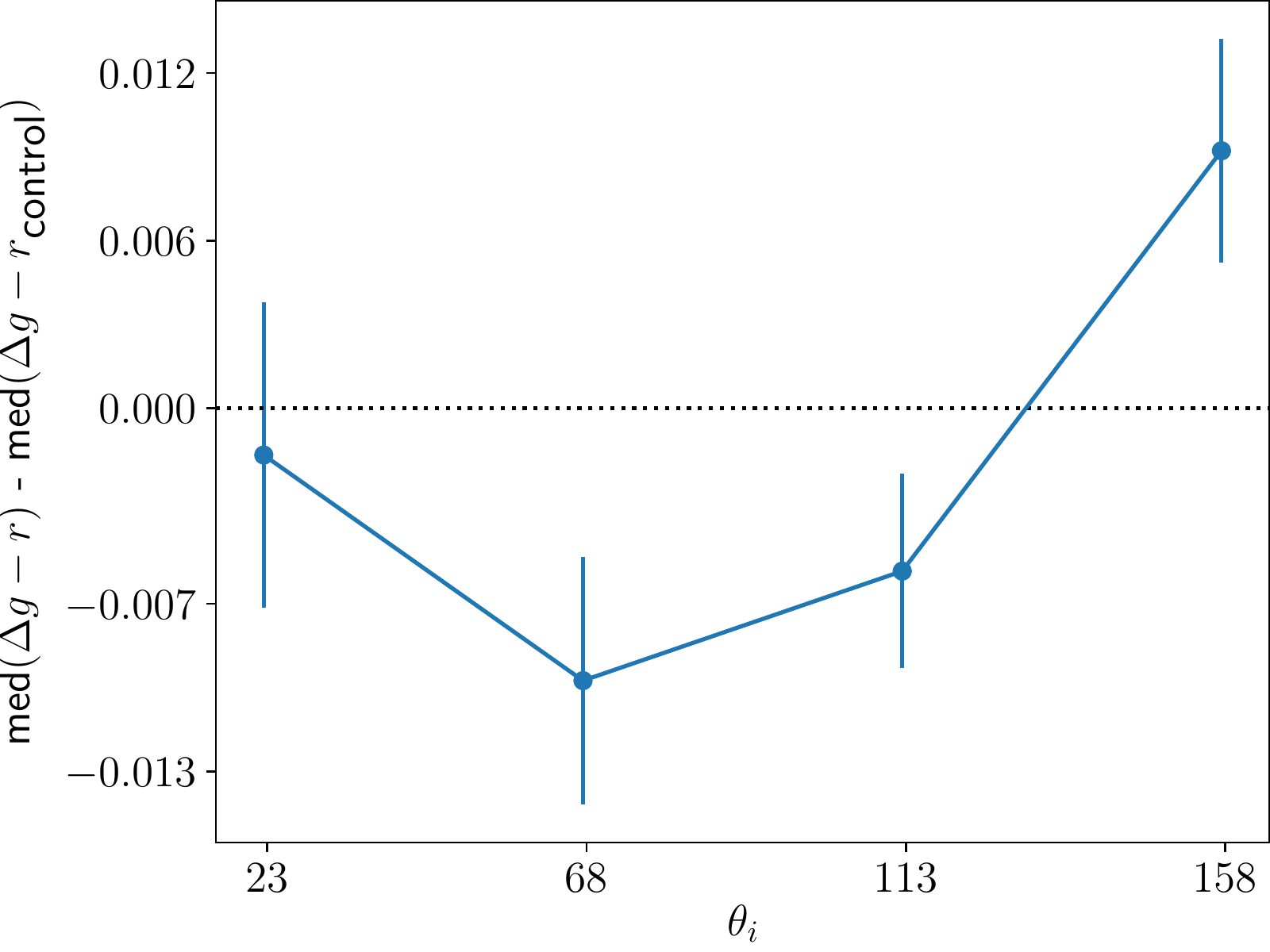}
		\includegraphics[width=.45\textwidth]{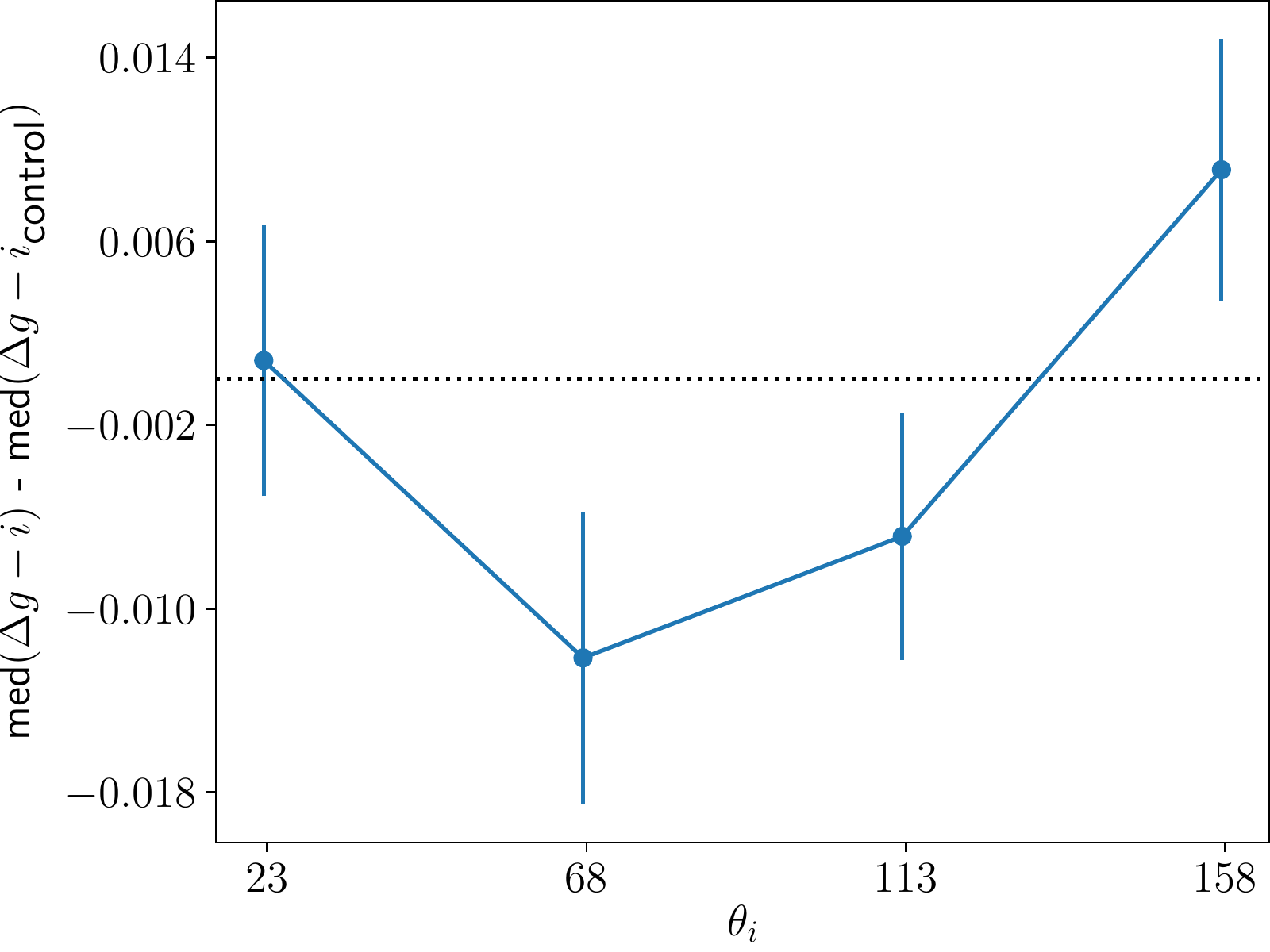}
		\caption{\label{fig_crumbs_lin} Difference between the average colour excess of background quasar as a function of the angle defined by the background galaxy, the galaxy in the group and the group centre, and the average colours of background galaxies in the control zones. LEFT: Average values for $g-r$ colour. RIGHT: Average values for $g-i$ colour.}
	\end{center}
\end{figure*}

We interpret these results as a consequence of a global reddening of background quasars by the intra-group medium plus an important contribution of material removed by in-falling galaxies. This hypothesis is reinforced by inspection to
Fig. \ref{fig_crumbs_lin} that shows the relation of the relative colour excess as a function of $\theta_i$ ($g-r$ in the left panel and $g-i$ in the right panel) subtracted to the colour excess of quasars associated to the control zones. 
It can be seen that both the $g-r$ and $g-i$ colour excess relative to the control sample increase towards the external radial direction ($\theta_i \sim 180$), as well as a marginal reddening towards the centre ($\theta_i \sim 0$). Thus, this analysis suggests that there is a contribution of ram pressure stripped dust in the infall region around groups.

Following, Equation 5 in \cite{mcgee2010}, we can estimate a dust mass density related to the colour excess measurements. Using the observed $g-r$ reddening of quasars at $\sim 180^{\circ}$ and assuming Milky Way dust physical properties as in \citep{mcgee2010}, we calculate a dust column density of $2034 \pm 877 M_{\sun}/kpc^2 h$. Taking into account the corresponding area under consideration (the area of the last bin), we obtain a dust mass of $5.8 \pm 2.5 \cdot 10^8 M_{\sun}/h$ (The median stellar mass of the sample is $10^{10.15} M_{\sun}$). We acknowledge that the estimated removed dust mass is quite high compared to the median galaxy stellar mass content by a factor of 2 or more \citep{calura17, kingfish}. However, we notice that a significant fraction of the stripped material can be associated to the extended gas reservoirs of galaxies, as well as to other galaxies in the same substructure.

\subsection{Dependence on group and galaxy parameters}

We have also explored the dependence of the observed trends of quasar colour excess and $\theta_i$ on different characteristics of group and member galaxies. For this aim, we have considered different sub-samples selected according to different properties of galaxies and groups.
These sub-samples take into account i) the angle between the vector to the centre of the group and \da semi-mayor axis with $\phi_s<45$ and $\phi_s>45$, ii) luminosity cuts  $M_r - 5 \log h < -19.5$ and $M_r - 5\log h > - 19.5$, iii) concentration cuts ($r_{90}/r_{50}<2.5$ and $r_{90}/r_{50}>2.5$), iv) the effects of the group halo mass by considering a threshold in $M_{\textmd{halo}} < 10^{13.9} M_{\sun}/h$ and $M_{\textmd{halo}} > 10^{13.9} M_{\sun}/h$, and v) galaxy colour. Given the relation between galaxy colour and luminosity we have not applied a single threshold colour cut, but use the median $g-r$ value as a function of absolute magnitude to separate the sample into red and blue galaxy sub-samples.

\begin{figure*}
	\begin{center}
		\includegraphics[width=.9\textwidth]{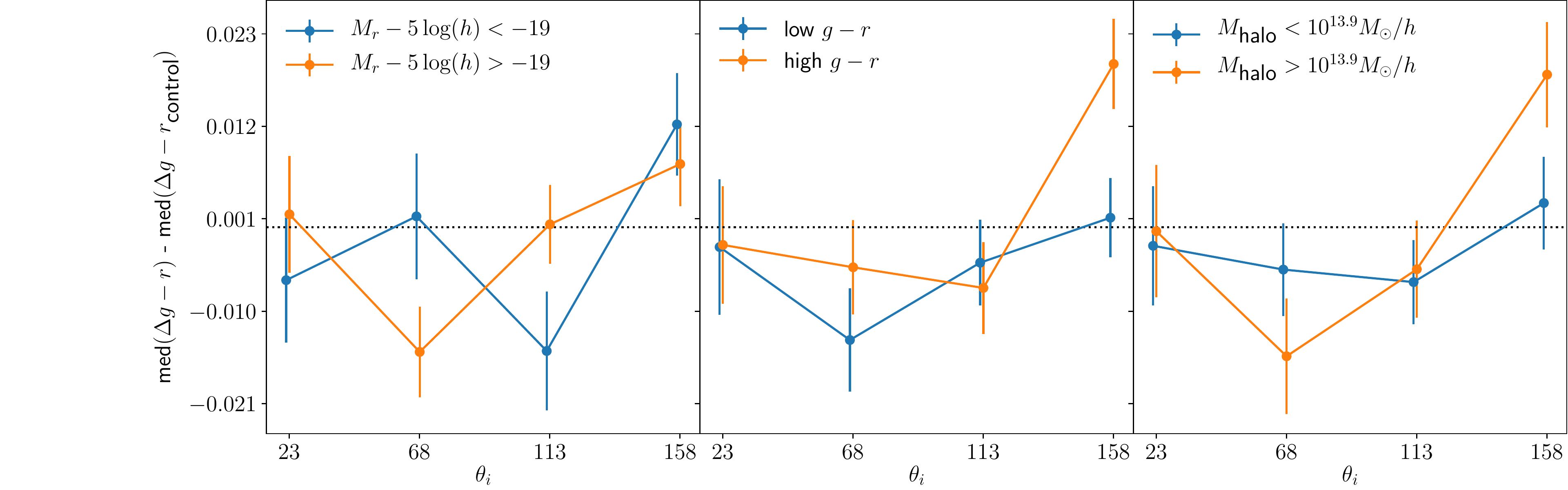}
		\caption{\label{fig_byStuff}Difference between the median of $g-r$ colour around galaxies, and the median colour in control zones. LEFT: Samples selected by absolute magnitude, MIDDLE: Samples selected by colour, RIGHT: Samples selected by group mass}
	\end{center}
\end{figure*}

All the sub-samples show the trend where background quasars with high $\theta_i$ tend to be redder in the presence of a galaxy.  However, there is a strong variation with the sub-sample luminosity, colour and group mass.
In the case of concentration and $\phi_s$ all the samples show a similar behaviour than the total sample, within uncertainties (determined via bootstrap). Fig. \ref{fig_byStuff} shows the results for the sub-samples divided by absolute magnitude of the galaxy, galaxy colour and group halo mass. As can be seen, the largest colour excess is associated to red, bright galaxies, and higher group mass, indicating larger amounts of removed dust.

These results are reasonable since bright galaxies have larger gas masses than the faint ones, and therefore more dust is likely to be stripped along their orbits in the groups. 
We also expect that galaxies suffering a stronger ram-pressure tend to loose higher amounts of material, in particular cold gas. Therefore star-formation should decrease and therefore redden their colours.
These effects can be seen in the middle panel of Fig. \ref{fig_byStuff}. We also observe that galaxies orbiting massive groups are likely to experience more stripping than galaxies in less massive groups with more tenous intragroup medium, as seen in the right panel of Fig. \ref{fig_byStuff}. We have tested the effects of the uncertainty of group centre positions due to different number of members, which could affect more strongly low mass groups. We find that using the same number of galaxies (n=4) to determine group centres for all group masses, the results remain unchanged.

\subsection{Internal effects on member galaxies}
\label{results_oranges}

\begin{figure}
	\begin{center}
		\includegraphics[width=.45\textwidth]{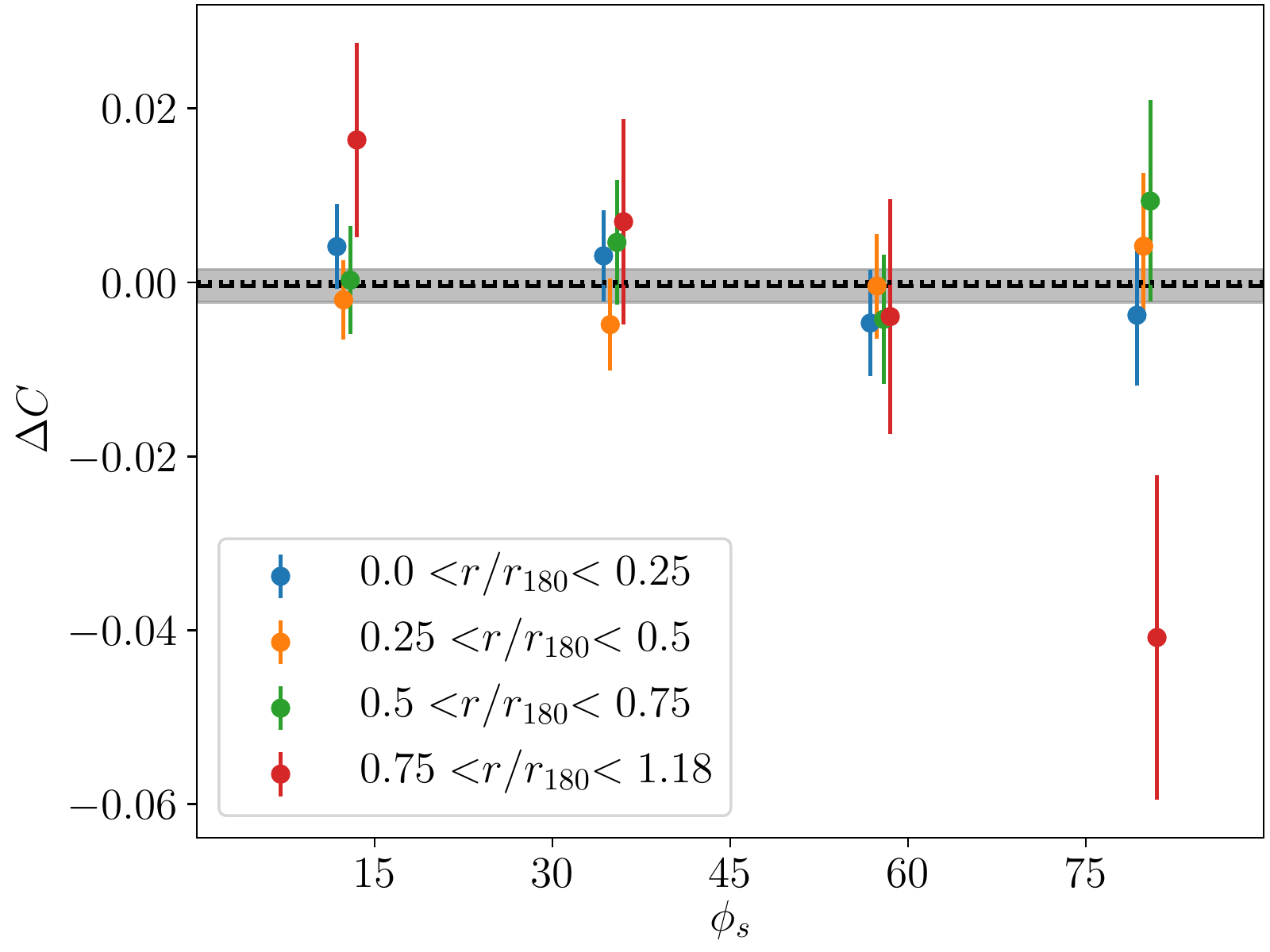}
		\caption{\label{fig_oranges} Differences between the colour of the leading and the trailing half as a function of the angle between the vector to the centre of the group and the semi-major axis of the galaxy, separated in subsamples according to the distance to the group centre in units of $r_{180}$. Dotted lines shows \da 0 value, while the dashed lines and the shadowed areas show the average value for all \da sample and the corresponding error.}
	\end{center}
\end{figure}

Following T16, we have computed the mean colour difference  between the leading and trailing galaxy halves $\Delta C$. Fig. \ref{fig_oranges} shows the behaviour of $\Delta C$ as a function of $\phi_s$ for subsamples selected according to the projected distance to the centre of \da groups in units of $r_{180}$. It is important to stress the fact that these results can only be compared with caution since the results in T16 are based on 3-D measurements, and consider the efficiency of star-formation of the gas particles instead of colour.

As can be seen, there are noticeable systematic differences between the mean colours of the leading and trailing halves for sub-samples of galaxies located beyond $0.75$  $r_{180}$ with high $\phi_s$, ie. galaxies whose disk tend to be perpendicular to the vector towards the group centre. Given their large group centric distance, it can be argued that \da vector to the centre is a suitable proxy for \da galaxy infall velocity vector. This is supported by the analysis of \citet{jaffe18} who find that galaxies in infall are preferentially located beyond 0.5 the group virial radius.

This result gives support to the idea that, under an appropriate setting, the leading half of galaxies residing in groups is bluer than the trailing half. This is in good agreement with the results of T16 and T19, where the leading half of galaxies in the simulation presents a higher SFR  due to the compression of the gas as the galaxy moves in the intracluster medium. This gas compression causes an increase of  pressure and density. This result is of small amplitude and no combination of other photometric colour bands ($u$, $g$, $r$ and $i$) provide larger differences than those observed in $g-r$. We notice that a stronger effect is expected in the $u-r$ colour index since it comprises the  4000 \r{A} break, often used as a proxy for star-formation activity \citep[e.g.][]{kauffmann03}. Yet,  photometric errors and low statistics, i.e. small extension of the galaxies make it not useful for this particular dataset.

\subsection{Exploring the relation between the effect on galaxies and removed debris}
\label{results_both}

\begin{figure}
	\begin{center}
		\includegraphics[width=.45\textwidth]{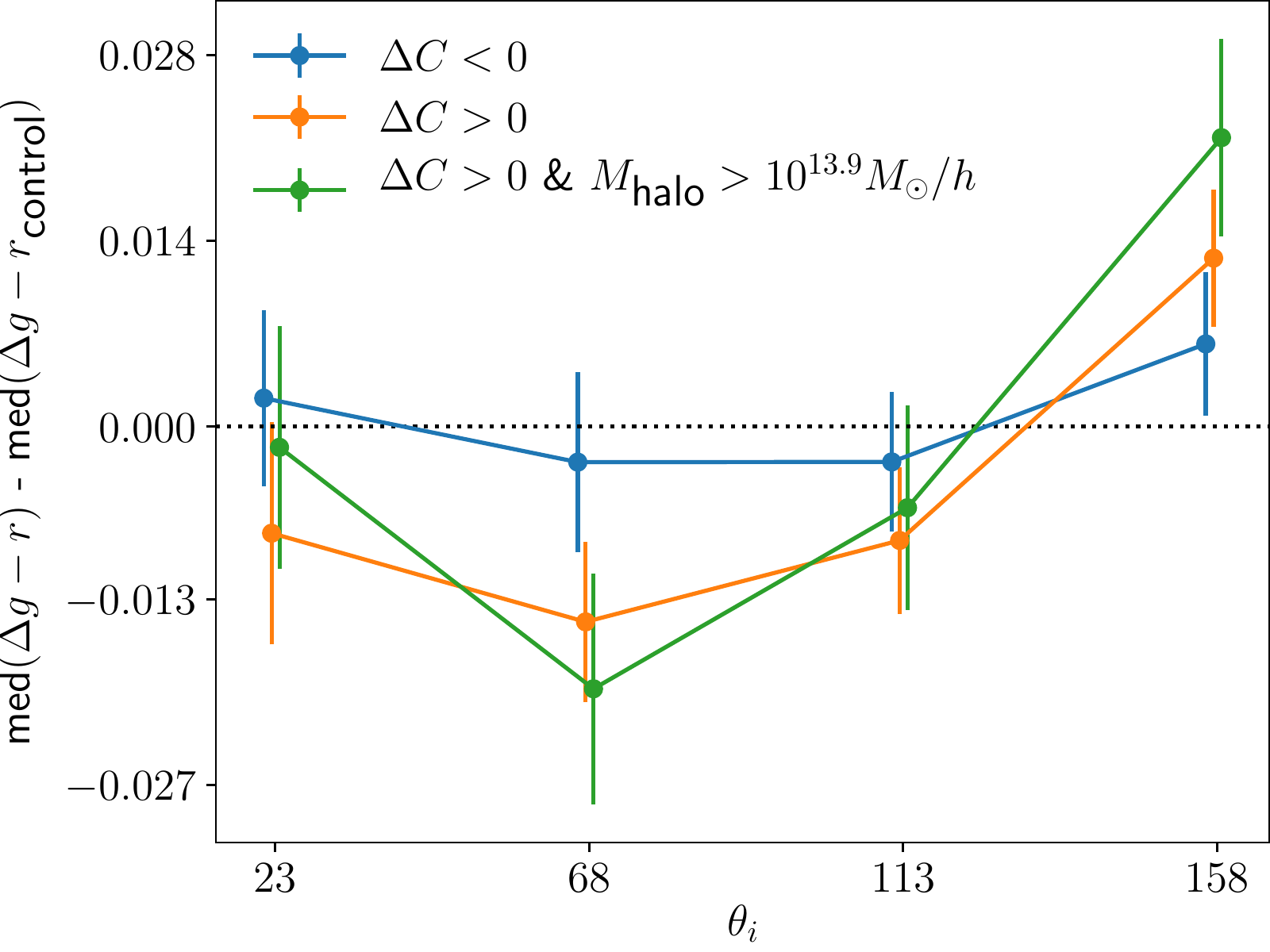}
		\caption{\label{fig_crumbs_byDif} $g-r$ colour excess of background quasars relative to the control sample  according to the difference of colour between the leading and trailing halves of the member galaxy.}
	\end{center}
\end{figure}

In this subsection we explore possible relations between the results obtained in  the two previous sub sections (\S \ref{results_crumbs} and \S \ref{results_oranges}). To this aim, we select galaxies with $r/r_{180}>0.5$ and separate this sample between those galaxies with $\Delta C$ larger than the mean (the leading half is bluer) and those with lower $\Delta C$ values (the trailing half is bluer). Fig \ref{fig_crumbs_byDif} shows the behaviour of the colour excess of the background quasars for these samples and the corresponding control zones.

It can be appreciated that there is a statistically significant difference between these two. Relative colour excess of quasars around galaxies with large $\Delta C$ values (ie. galaxies whose leading halves are redder than their corresponding trailing halves) have a much stronger signal compared to the low $\Delta C$ subsample. This signal is even stronger for galaxies with high $\Delta C$ located in massive halos ($M_{halo} > 10^{13.9} M_{\sun}h^{-1}$, green dots).

It can be argued that galaxies which have experienced strong stripping due to ram pressure are those associated to the strongest removal of dust. These more efficently stripped galaxies have a weaker gravitational potential and their leading regions should experience gas stripping rather than gas compression. Another explanation is that most massive haloes are more efficient stripping the gas as it is observed in Fig. \ref{fig_crumbs_byDif} (green dots). This scenario could explain why galaxies with a leading half with a lower star-formation exhibit signs of removed gas as measured by the background quasar colour excess. Therefore, we interpret the previous result as a further evidence that the same astrophysical processes, ram-pressure, acting on galaxies making them asymmetric, also generates the quasar reddening which we associate to dust removed from their interstellar medium.

\section{Discussion and conclusions}
\label{conclusions}

In this paper we have studied the presence of dust around galaxies in groups in the local universe. The procedures were based on  the systematic reddening of background quasars in regions neighbours of groups. We carefully take into account the redshift and angular position dependence of the observed mean quasar colours, which allow us to derive more precise mean colour excesses around group galaxies. We have focused on disc galaxies observed perpendicular to the plane of the sky, and at distances to group centers consistent with the infall regions of groups.
Assuming  dust properties consistent with those of the Milky Way, dust masses of $5.8 \pm 2.5 \cdot 10^8 M_{\sun}/h$ per galaxy are inferred, implying large fraction of dust stripped from galaxies orbiting groups.

We stress the fact that significant reddening of background quasars are derived for the subsample of the most massive groups, $M_{\textmd{halo}}>10^{13.9} M_{\sun}/h$, as expected under the assumption of ram pressure stripping by the intragroup medium as the cause for the stripped dust.

A photometric study  of the galaxies is performed to derive a colour asymmetry parameter relative to the group centre direction from the galaxy position. This analysis shows that regions of galaxies facing the centre are bluer than their opposite counterparts, a fact that we interpret in terms of the effects of gas compression and star-formation along the galaxy motion towards the group centre.   

Consistently, galaxies with the smallest colour asymmetries show the largest amounts of dust radially outwards compared to a control sample with no galaxy at the same group-centric distance. In control regions, it is certainly possible that the colour excess measurements come from a correlated filament environment hosting infalling galaxies. This issue will be studied in more detail in a forthcoming paper.

We conclude that dust removal is very efficient in galaxies infalling onto groups, particularly onto the most massive ones. The fact that galaxies with the smallest group-centric colour asymmetries are associated to the strongest reddening of background quasars suggests that gas dynamics and removal induced by ram pressure from the intragroup medium are suitable mechanisms acting on the leading and trailing regions of galaxies.

We notice that other mechanisms aside from ram pressure may be present in infalling galaxies. For instance, gas expelled by supernovae winds isotropically from galaxies likely to be destroyed by the radiation from the hot ICM towards groups centres, creating the observed reddening trends. However, the internal radial colour asymmetry and its correlation to the background quasar colour excess around galaxies with redder leading halves provides a hint that ram pressure may be the most appropriate mechanism to explain the observations.

\section*{ACKNOWLEDGEMENTS}
This work was supported by Consejo Nacional de Investigaciones
Cient\'ificas y T\'ecnicas (CONICET).
PTI acknowledges funding from ANILLO-ACT-1417 and Pontificia Universidad Cat\'olica for hosting her during the creation of this work.  NP acknowledges support by BASAL CATA AFB-170002 and Fondecyt Regular 1191813.

This research made use of Astropy,\footnote{http://www.astropy.org} a community-developed core Python package for Astronomy \citep{astropy18}, Numpy \citep{numpy}, Scipy \citep{scipy} and Matplotlib \citep{matplotlib}.

Funding for the Sloan Digital Sky Survey IV has been provided by the Alfred P. Sloan Foundation, the U.S. Department of Energy Office of Science, and the Participating Institutions. SDSS-IV acknowledges
support and resources from the Center for High-Performance Computing at
the University of Utah. The SDSS web site is www.sdss.org.

SDSS-IV is managed by the Astrophysical Research Consortium for the 
Participating Institutions of the SDSS Collaboration including the 
Brazilian Participation Group, the Carnegie Institution for Science, 
Carnegie Mellon University, the Chilean Participation Group, the French Participation Group, Harvard-Smithsonian Center for Astrophysics, 
Instituto de Astrof\'isica de Canarias, The Johns Hopkins University, Kavli Institute for the Physics and Mathematics of the Universe (IPMU) / 
University of Tokyo, the Korean Participation Group, Lawrence Berkeley National Laboratory, 
Leibniz Institut f\"ur Astrophysik Potsdam (AIP),  
Max-Planck-Institut f\"ur Astronomie (MPIA Heidelberg), 
Max-Planck-Institut f\"ur Astrophysik (MPA Garching), 
Max-Planck-Institut f\"ur Extraterrestrische Physik (MPE), 
National Astronomical Observatories of China, New Mexico State University, 
New York University, University of Notre Dame, 
Observat\'ario Nacional / MCTI, The Ohio State University, 
Pennsylvania State University, Shanghai Astronomical Observatory, 
United Kingdom Participation Group,
Universidad Nacional Aut\'onoma de M\'exico, University of Arizona, 
University of Colorado Boulder, University of Oxford, University of Portsmouth, 
University of Utah, University of Virginia, University of Washington, University of Wisconsin, 
Vanderbilt University, and Yale University.
 
\bibliographystyle{mnras}
\footnotesize
\bibliography{bib2.bib}

\begin{thebibliography}{}
\makeatletter
\relax
\def\mn@urlcharsother{\let\do\@makeother \do\$\do\&\do\#\do\^\do\_\do\%\do\~}
\def\mn@doi{\begingroup\mn@urlcharsother \@ifnextchar [ {\mn@doi@}
  {\mn@doi@[]}}
\def\mn@doi@[#1]#2{\def\@tempa{#1}\ifx\@tempa\@empty \href
  {http://dx.doi.org/#2} {doi:#2}\else \href {http://dx.doi.org/#2} {#1}\fi
  \endgroup}
\def\mn@eprint#1#2{\mn@eprint@#1:#2::\@nil}
\def\mn@eprint@arXiv#1{\href {http://arxiv.org/abs/#1} {{\tt arXiv:#1}}}
\def\mn@eprint@dblp#1{\href {http://dblp.uni-trier.de/rec/bibtex/#1.xml}
  {dblp:#1}}
\def\mn@eprint@#1:#2:#3:#4\@nil{\def\@tempa {#1}\def\@tempb {#2}\def\@tempc
  {#3}\ifx \@tempc \@empty \let \@tempc \@tempb \let \@tempb \@tempa \fi \ifx
  \@tempb \@empty \def\@tempb {arXiv}\fi \@ifundefined
  {mn@eprint@\@tempb}{\@tempb:\@tempc}{\expandafter \expandafter \csname
  mn@eprint@\@tempb\endcsname \expandafter{\@tempc}}}

\bibitem[\protect\citeauthoryear{{Abazajian} et~al.,}{{Abazajian}
  et~al.}{2009}]{dr7}
{Abazajian} K.~N.,  et~al., 2009, \mn@doi [\apjs]
  {10.1088/0067-0049/182/2/543}, \href
  {http://adsabs.harvard.edu/abs/2009ApJS..182..543A} {182, 543}

\bibitem[\protect\citeauthoryear{Aihara et~al.,}{Aihara et~al.}{2011}]{dr8}
Aihara H.,  et~al., 2011, \mn@doi [\apjs] {10.1088/0067-0049/193/2/29}, 193, 29

\bibitem[\protect\citeauthoryear{{Astropy Collaboration} et~al.,}{{Astropy
  Collaboration} et~al.}{2018}]{astropy18}
{Astropy Collaboration} et~al., 2018, \mn@doi [\aj] {10.3847/1538-3881/aabc4f},
  \href {https://ui.adsabs.harvard.edu/abs/2018AJ....156..123A} {156, 123}

\bibitem[\protect\citeauthoryear{{Barnes} \& {Hernquist}}{{Barnes} \&
  {Hernquist}}{1996}]{barnes96}
{Barnes} J.~E.,  {Hernquist} L.,  1996, \mn@doi [\apj] {10.1086/177957}, \href
  {https://ui.adsabs.harvard.edu/abs/1996ApJ...471..115B} {471, 115}

\bibitem[\protect\citeauthoryear{{Boselli} \& {Gavazzi}}{{Boselli} \&
  {Gavazzi}}{2006}]{boselli06}
{Boselli} A.,  {Gavazzi} G.,  2006, \mn@doi [\pasp] {10.1086/500691}, \href
  {https://ui.adsabs.harvard.edu/abs/2006PASP..118..517B} {118, 517}

\bibitem[\protect\citeauthoryear{{Calura} et~al.,}{{Calura}
  et~al.}{2017}]{calura17}
{Calura} F.,  et~al., 2017, \mn@doi [\mnras] {10.1093/mnras/stw2749}, \href
  {https://ui.adsabs.harvard.edu/abs/2017MNRAS.465...54C} {465, 54}

\bibitem[\protect\citeauthoryear{Crain et~al.,}{Crain et~al.}{2015}]{eagle2}
Crain R.~A.,  et~al., 2015, \mn@doi [\mnras] {10.1093/mnras/stv725}, 450, 1937

\bibitem[\protect\citeauthoryear{{Dale} et~al.,}{{Dale}
  et~al.}{2017}]{kingfish}
{Dale} D.~A.,  et~al., 2017, \mn@doi [\apj] {10.3847/1538-4357/aa6032}, \href
  {https://ui.adsabs.harvard.edu/abs/2017ApJ...837...90D} {837, 90}

\bibitem[\protect\citeauthoryear{{De Lucia}, {Kauffmann}  \& {White}}{{De
  Lucia} et~al.}{2004}]{delucia04}
{De Lucia} G.,  {Kauffmann} G.,   {White} S. D.~M.,  2004, \mn@doi [\mnras]
  {10.1111/j.1365-2966.2004.07584.x}, \href
  {https://ui.adsabs.harvard.edu/abs/2004MNRAS.349.1101D} {349, 1101}

\bibitem[\protect\citeauthoryear{{Gunn} \& {Gott}}{{Gunn} \&
  {Gott}}{1972}]{gunn72}
{Gunn} J.~E.,  {Gott} J.~Richard I.,  1972, \mn@doi [\apj] {10.1086/151605},
  \href {https://ui.adsabs.harvard.edu/abs/1972ApJ...176....1G} {176, 1}

\bibitem[\protect\citeauthoryear{{Guo} et~al.,}{{Guo} et~al.}{2011}]{guo11}
{Guo} Q.,  et~al., 2011, \mn@doi [\mnras] {10.1111/j.1365-2966.2010.18114.x},
  \href {https://ui.adsabs.harvard.edu/abs/2011MNRAS.413..101G} {413, 101}

\bibitem[\protect\citeauthoryear{Hunter}{Hunter}{2007}]{matplotlib}
Hunter J.~D.,  2007, \mn@doi [Computing in Science \& Engineering]
  {10.1109/MCSE.2007.55}, 9, 90

\bibitem[\protect\citeauthoryear{{Jaff{\'e}} et~al.,}{{Jaff{\'e}}
  et~al.}{2018}]{jaffe18}
{Jaff{\'e}} Y.~L.,  et~al., 2018, \mn@doi [\mnras] {10.1093/mnras/sty500},
  \href {http://adsabs.harvard.edu/abs/2018MNRAS.476.4753J} {476, 4753}

\bibitem[\protect\citeauthoryear{Jones, Oliphant, Peterson  et~al.}{Jones
  et~al.}{2001}]{scipy}
Jones E.,  Oliphant T.,  Peterson P.,   et~al., 2001, {{SciPy}: Open source
  scientific tools for {Python}}, \url {http://www.scipy.org/}

\bibitem[\protect\citeauthoryear{{Kauffmann} et~al.,}{{Kauffmann}
  et~al.}{2003}]{kauffmann03}
{Kauffmann} G.,  et~al., 2003, \mn@doi [\mnras]
  {10.1111/j.1365-2966.2003.07154.x}, \href
  {http://adsabs.harvard.edu/abs/2003MNRAS.346.1055K} {346, 1055}

\bibitem[\protect\citeauthoryear{{May}}{{May}}{2007}]{may07}
{May} B.~H.,  2007, {A Survey of Radial Velocities in the Zodiacal Dust Cloud},
  \mn@doi{10.1007/978-0-387-77706-1.
}

\bibitem[\protect\citeauthoryear{{McGee} \& {Balogh}}{{McGee} \&
  {Balogh}}{2010}]{mcgee2010}
{McGee} S.~L.,  {Balogh} M.~L.,  2010, \mn@doi [\mnras]
  {10.1111/j.1365-2966.2010.16616.x}, \href
  {https://ui.adsabs.harvard.edu/\#abs/2010MNRAS.405.2069M} {405, 2069}

\bibitem[\protect\citeauthoryear{{Moore}, {Katz}, {Lake}, {Dressler}  \&
  {Oemler}}{{Moore} et~al.}{1996}]{moore96}
{Moore} B.,  {Katz} N.,  {Lake} G.,  {Dressler} A.,   {Oemler} A.,  1996,
  \mn@doi [\nat] {10.1038/379613a0}, \href
  {https://ui.adsabs.harvard.edu/abs/1996Natur.379..613M} {379, 613}

\bibitem[\protect\citeauthoryear{{Oke} \& {Sandage}}{{Oke} \&
  {Sandage}}{1968}]{oke1968}
{Oke} J.~B.,  {Sandage} A.,  1968, \mn@doi [\apj] {10.1086/149737}, \href
  {https://ui.adsabs.harvard.edu/\#abs/1968ApJ...154...21O} {154, 21}

\bibitem[\protect\citeauthoryear{{P{\^a}ris} et~al.,}{{P{\^a}ris}
  et~al.}{2017}]{BOSS}
{P{\^a}ris} I.,  et~al., 2017, \mn@doi [\aap] {10.1051/0004-6361/201527999},
  \href {https://ui.adsabs.harvard.edu/\#abs/2017A\&A...597A..79P} {597, A79}

\bibitem[\protect\citeauthoryear{{Poggianti} et~al.,}{{Poggianti}
  et~al.}{2016}]{Poggianti2016}
{Poggianti} B.~M.,  et~al., 2016, VizieR Online Data Catalog, \href
  {http://adsabs.harvard.edu/abs/2016yCat..51510078P} {515}

\bibitem[\protect\citeauthoryear{Schaye et~al.,}{Schaye et~al.}{2015}]{eagle1}
Schaye J.,  et~al., 2015, \mn@doi [\mnras] {10.1093/mnras/stu2058}, 446, 521

\bibitem[\protect\citeauthoryear{{Schlegel}, {Finkbeiner}  \&
  {Davis}}{{Schlegel} et~al.}{1998}]{schlegel98}
{Schlegel} D.~J.,  {Finkbeiner} D.~P.,   {Davis} M.,  1998, \mn@doi [\apj]
  {10.1086/305772}, \href
  {https://ui.adsabs.harvard.edu/abs/1998ApJ...500..525S} {500, 525}

\bibitem[\protect\citeauthoryear{{Steinhauser}, {Schindler}  \&
  {Springel}}{{Steinhauser} et~al.}{2016}]{steinhauser16}
{Steinhauser} D.,  {Schindler} S.,   {Springel} V.,  2016, \mn@doi [\aap]
  {10.1051/0004-6361/201527705}, \href
  {https://ui.adsabs.harvard.edu/abs/2016A\&A...591A..51S} {591, A51}

\bibitem[\protect\citeauthoryear{{Tecce}, {Cora}, {Tissera}, {Abadi}  \&
  {Lagos}}{{Tecce} et~al.}{2010}]{tecce10}
{Tecce} T.~E.,  {Cora} S.~A.,  {Tissera} P.~B.,  {Abadi} M.~G.,   {Lagos} C.
  D.~P.,  2010, \mn@doi [\mnras] {10.1111/j.1365-2966.2010.17262.x}, \href
  {https://ui.adsabs.harvard.edu/abs/2010MNRAS.408.2008T} {408, 2008}

\bibitem[\protect\citeauthoryear{{Troncoso Iribarren}, {Padilla}, {Contreras},
  {Rodriguez}, {Garc\'{\i}a-Lambas}  \& {Lagos}}{{Troncoso Iribarren}
  et~al.}{2016}]{TI16}
{Troncoso Iribarren} P.,  {Padilla} N.,  {Contreras} S.,  {Rodriguez} S.,
  {Garc\'{\i}a-Lambas} D.,   {Lagos} C.,  2016, \mn@doi [Galaxies]
  {10.3390/galaxies4040077}, \href
  {http://adsabs.harvard.edu/abs/2016Galax...4...77T} {4, 77}

\bibitem[\protect\citeauthoryear{Yang, Mo, van~den Bosch, Pasquali, Li  \&
  Barden}{Yang et~al.}{2007}]{yang2007}
Yang X.,  Mo H.~J.,  van~den Bosch F.~C.,  Pasquali A.,  Li C.,   Barden M.,
  2007, \mn@doi [\apj] {10.1086/522027}, 671, 153

\bibitem[\protect\citeauthoryear{Yang, Mo, van~den Bosch, Zhang  \& Han}{Yang
  et~al.}{2012}]{yang2012}
Yang X.,  Mo H.~J.,  van~den Bosch F.~C.,  Zhang Y.,   Han J.,  2012, \mn@doi
  [\apj] {10.1088/0004-637X/752/1/41}, 752, 41

\bibitem[\protect\citeauthoryear{{York} et~al.,}{{York} et~al.}{2000}]{sdss}
{York} D.~G.,  et~al., 2000, \mn@doi [\aj] {10.1086/301513}, 120, 1579

\bibitem[\protect\citeauthoryear{{van der Walt}, {Colbert}  \&
  {Varoquaux}}{{van der Walt} et~al.}{2011}]{numpy}
{van der Walt} S.,  {Colbert} S.~C.,   {Varoquaux} G.,  2011, \mn@doi
  [Computing in Science and Engineering] {10.1109/MCSE.2011.37}, \href
  {https://ui.adsabs.harvard.edu/abs/2011CSE....13b..22V} {13, 22}

\makeatother
\end{thebibliography}

\label{lastpage}

\end{document}